# Disorder-Driven Spin-Reorientation in Multiferroic $h$-$YMn_{1-x}Fe_xO_3$

Sonu Namdeo[1], S.S.S. Rao[1], S.D. Kaushik[2], V. Siruguri[2], and A.M. Awasthi[1*]

[1] *UGC-DAE-Consortium for Scientific Research, University Campus, Khandwa Road, Indore 452 001, India.*

[2] *UGC-DAE-Consortium for Scientific Research, Mumbai Centre, R5 Shed, Bhabha Atomic Research Centre, Mumbai 400 085, India*

**ABSTRACT**

Magnetic structure evolution of multiferroic hexagonal $YMn_{1-x}Fe_xO_3$ ($x$= 0, 0.05, and 0.1) has been studied by carrying out detailed temperature-dependent neutron diffraction at zero and 5T fields. Thermodynamic data confirm antiferromagnetic ordering at $T_N$ in all the compositions. Our sub-$T_N$ neutron diffraction results assign the magnetic structure of pure $YMnO_3$ to $\Gamma_1$ irreducible representation. Over the perturbative-doping range, magnetic structure changes via $\Gamma_1$+$\Gamma_2$ for $YMn_{0.95}Fe_{0.05}O_3$ on to $\Gamma_2$ for $YMn_{0.9}Fe_{0.1}O_3$, as the maiden compositional analogue of spin-reorientation; its occurrence in temperature-domain already reported for several manganites. Moreover, while the large thermal isostructural changes observed above $T_N$ are subdued in the ordered state, small alterations by the applied 5T-field are relatively uniform across, confirming strong magneto-elastic nature of the system. Decrease of the ordered magnetic moment ($\mu_{ord}$) and planar magnetic frustration noted with Fe-doping is enhanced by the applied field; apparently through canting.

[*] Corresponding Author. E-mail: amawasthi@csr.res.in. Tel.: +91 731 2763913. Fax: +91 731 2762294.

Materials that possess coexistent magnetic and electrical orderings in single phase are known as Multiferroics.[1] Hexagonal $YMnO_3$ is a member of their special class known as geometrically frustrated multiferroics, and is one of the most intensively studied $h$-$R$MnO$_3$.[2] It undergoes ferroelectric (FE) transition at $T_C \sim 900$K and antiferromagnetic (AFM) ordering at $T_N \sim$70K.[3-4] The crystal structure of $h$-YMnO$_3$ consists of alternating layers of $MnO_5$ trigonal bipyramids. These bipyramids (in trimer-combinations, sharing a common planar Oxygen) are corner-linked in the *ab*-plane to form a triangular lattice, and are separated from one another along the *c*-axis by the planes of Y-ions.[5] The geometric effect of buckling of the trimerized MnO5-bipyramids is accompanied by the inversion-symmetry-breaking displacements of Y-layers, originating the ferroelectric polarization.[6] Regarding the magnetic structure, below $T_N$ the $Mn^{3+}$ moments order in the *a-b* plane, and form a 2D non-collinear spin structure, characterized by an angle of 120$^o$ between the neighbouring spins.[7-9] Consequent to the 2D magnetism and triangular lattice, the magnetic structure of $YMnO_3$ is frustrated, characterized by the frustration factor $f = |\Theta_{CW}|/T_N$, found to be as large as 7.8 for $YMnO_3$.[10] Here, $\Theta_{CW}$ is the (negative) intercept on the temperature-axis of the (interpolated) high-temperature $T$-linear (Curie-Weiss) inverse susceptibility of the independent-spins. The antiferromagnetic (AFM) order is governed by the in-plane Mn-O-Mn super-exchange interaction, whereas the Mn-O-O-Mn super-super-exchange interaction between the stacked triangular planes is about two order of magnitude weaker. This weak inter-layer ($z$=0 and $z$=1/2),[4,7,11] configuration is the one that differentiates among the several magnetic structures proposed for $YMnO_3$, and needs to be precisely investigated. Although the magnetic structure of $YMnO_3$ has been studied for years, the nature of the coupling between the $z$=0 and $z$=1/2 planes is still unsettled.

According to group theory analysis, altogether six magnetic structures are found to be possible with $\boldsymbol{k}$ = 0 vector (Goldstone modes); four 1D ($\Gamma_1$, $\Gamma_2$, $\Gamma_3$, and $\Gamma_4$) and two 2D ($\Gamma_5$ and $\Gamma_6$) irreducible representations.[7] However, all hexagonal manganites are known to particularly favour the 1D irreducible representations. Out of these four, $\Gamma_1$ and $\Gamma_3$ have antiparallel coupling of $z$=0 and $z$=1/2 moments, while $\Gamma_2$ and $\Gamma_4$ have parallel coupling. Moreover, in $\Gamma_1$ and $\Gamma_4$ representations, moments are perpendicular to the [100] axis, while in $\Gamma_2$ and $\Gamma_3$, moments are parallel to the [100] axis. The magnetic structure of $h$-$R$MnO$_3$ was first studied in the early 1960's by the group of Bertaut and co-workers.[4] According to them, the neutron-diffraction data of $YMnO_3$ can be explained by ($\Gamma_1$+$\Gamma_3$) representation. On the other hand, Chandrasekhar *et al.*[12] found $\Gamma_1$ representation of $YMnO_3$, while Munoz *et.al.* [7] and Park *et al*[8] found that the magnetic peaks of $YMnO_3$ can be explained by either $\Gamma_1$ or $\Gamma_3$ representations. Further, it was recently suggested that even pure $h$-YMnO$_3$ might have a mixed magnetic structure, with a mixing coefficient 0.19% of $\Gamma_1/\Gamma_3$ and $\Gamma_2/\Gamma_4$.[13] Furthermore, upon doping nonmagnetic atoms (c.f., Al, Ru, and Zn) only up to 10% at the Mn-site, mixing of the two representations gets increased.[14] Therefore, the magnetic structure of $YMnO_3$ remains somewhat uncertain.

Recently, Sharma *et al.*[15] carried out a comparative study by $Ti^{4+}$ ($d^0$), $Fe^{3+}$ ($d^5$), and $Ga^{3+}$ ($d^{10}$) (only 10%) doping at the Mn-site. Below $T_N$ ~75K, the magnetic structure of $YMnO_3$ was described as a linear combination of irreducible representations $\Gamma_3$ and $\Gamma_4$, the ratio of which changed with lowering the temperature. The change of magnetic structure to $\Gamma_2$ was caused by Ti, and with Fe-doping, the magnetic structure was described by the irreducible representation $\Gamma_3$, which below 35K underwent spin-reorientation to $\Gamma_3$ plus 51% $\Gamma_4$. On the other hand, rare-earth (Er) doping at Y-site[12] resulted in the change of magnetic ground state from $\Gamma_1$ to $\Gamma_2$, and also induced spin-reorientation transition via intermediate ($\Gamma_1$+$\Gamma_2$) configuration for half-doped $Y_{0.5}Er_{0.5}MnO_3$. Kozlenko *et.al.*[16] in their pressure-dependent neutron diffraction study on $YMnO_3$ showed that on application of up to 6GPa external pressure, pure $YMnO_3$ also undergoes spin-reorientation transition, analogous to temperature domain ($T_{sr}$)[7,17] in $ScMnO_3$ and $HoMnO_3$. Lee *et al.*[18] found via high-resolution neutron diffraction that all structural parameters abruptly change at $T_N$; evidencing strong and direct spin-lattice coupling. Moreover, recently giant magneto-elastic (MEL) coupling has been observed in $YMnO_3$,[19-20] and the dominance of this MEL over a weaker magneto-electric (ME) coupling in the coexistent FE and AFM phases of $YMnO_3$ has been reported by field-dependent neutron diffraction experiments.[21] The effects of magnetic doping on the magnetic structure of $YMnO_3$ have rarely been investigated. Therefore, perturbed by the magnetic Fe-ion at the Mn-site, we have explored the modification of the magnetic structure of $YMnO_3$. This study provides valuable new insights to correlate the changes of magnetic structure with the properties.

Polycrystalline samples of $YMn_{1-x}Fe_xO_3$ ($x$ = 0, 0.05, and 0.1) were synthesized by conventional solid state reaction route. Stoichiometric amounts of high purity (> 99.99%) $Y_2O_3$, $MnO_2$, and $Fe_2O_3$ powders were thoroughly mixed and calcined at 1200$^o$C for 27 hrs. The resultant powders were pelletized and sintered in air at 1400$^o$C for a total of 27 hrs, along with intermediate grinding. XRD of the prepared samples were obtained by Bruker D8 X-ray diffractrometer, with Cu-$K_\alpha$ radiation ($\lambda$=1.54Å). Magnetization and specific heat data were taken using a Quantum Design 14T MPMS-Vibrating sample magnetometer (VSM) and PPMS, respectively. The neutron diffraction (ND) measurements were carried out at wavelength of 1.48Å on powder samples, using the multi-position-sensitive-detector-based focusing crystal diffractometer, set up by UGC-DAE CSR (Mumbai Centre), at the National Facility for Neutron Beam Research (NFNBR), Dhruva reactor, Mumbai (India). ND patterns were collected at 10K, 30K, 50K, 60K, 70K, 100K, 200K, and 300K. The samples were placed in Vanadium-can, which was directly exposed to the neutron beam at 300K, while for low-temperature measurements, Vanadium-can filled with powder samples was loaded in a Cryogenic-make cryogen-free magnet system. ND data were taken in the warming cycle and were Rietveld-refined for both crystalline and magnetic parts, using FULLPROF program.[22]

The as-prepared samples were first subjected to XRD measurements for its structural characterization. All the XRD patterns were indexed to hexagonal structure with space group $P6_3cm$. Figure 1 shows the specific heat in the temperature range from 2-200K under zero and 5T magnetic fields. The $C_p$-data show a prominent peak, confirming the AFM ordering temperature which well-correlates with our magnetization data, discussed below. The absence of any secondary peak around 43K in the data confirms that as-prepared samples do not contain any impurity phase (e.g., ferrimagnetic $Mn_3O_4$).[23] Also, the $T$-position of this peak is unaffected up to 5T magnetic field. As the well-known feature of these hexagonal manganites, signature of magnetic frustration in the $C_p$-$T$ data have been reported[24] as large magnetic-entropy reductions above $T_N$.

Before performing neutron diffraction (ND), we measured DC magnetization from 10 to 300K, in an applied field of 100 Oe under ZFC and FC conditions. All the specimens exhibit very weak antiferromagnetic (AFM) transition as small kinks at $T_N$, shown in $1/\chi = H/M$ vs. $T$ plots (fig.2, left panels). For robust AFM ordering (no magnetic-frustration), spin-correlations above the transition temperature are absent, and the high-temperature linear Curie-Weiss behaviour ($1/\chi \sim [T\text{-}\Theta_{CW}]$) of independent (i.e., uncorrelated) spins continues down to $T_N$. In such cases, magnitude of the Curie-Weiss temperature $|\Theta_{CW}|$ exactly matches with that of the ordering temperature $T_N$ (frustration-parameter $f = |\Theta_{CW}|/T_N \equiv 1$). By the same token, any deviation (above $T_N$) from the high-$T$ linear-behaviour is taken as indication of magnetic frustration, concurrent with $f > 1$.[25-26] Magnetic frustration is evident in the present $1/\chi$-$T$ data as well; deviating from the Curie-Weiss linear-fit at $T \geq$ 150K > $T_N$. The interpolated linear fits over 150-300K to the ($1/\chi$-$T$) plot intercept on the (-ve) temperature axis, providing the value of Curie-Weiss temperature ($\Theta_{CW}$) and the fit's-slope provides effective magnetic moment ($\mu_{eff}$). The value of $\Theta_{CW}$ (estimated error ±5K) is found to decrease from $\Theta_{CW}$ = -533.59K for $YMnO_3$ to -463.88K for $YMn_{0.95}Fe_{0.05}O_3$, and to -426.75K for $YMn_{0.90}Fe_{0.10}O_3$. Since $\Theta_{CW}$ is a measure of the AFM coupling strength between the Mn-ions, the results suggest that Fe-doping suppresses the Mn-O-Mn AFM superexchange. Simultaneously, the estimated effective magnetic moment (estimated error $\pm 0.05\mu_B$) also shows a similar change: $\mu_{eff} = 5.45\mu_B$ estimated for $YMnO_3$ reduces to $5.27\mu_B$ for $YMn_{0.95}Fe_{0.05}O_3$ and to $5.04\mu_B$ for $YMn_{0.9}Fe_{0.1}O_3$. As $Fe^{+3}$ has higher magnetic moment ($5.92\mu_B$) compared to that of $Mn^{+3}$ ($4.90\mu_B$), the observed decreasing trend of effective magnetic moment rules out that Fe-ions are in the $Fe^{+3}$ state, which is consistent with a previous report by some of us.[27] In addition, the frustration parameter $f = |\Theta_{CW}|/T_N$ decreases with Fe-doping from 7.51 for $YMnO_3$ to 6.53 for $YMn_{0.95}Fe_{0.05}O_3$, on to 6.10 for $YMn_{0.9}Fe_{0.1}O_3$. This trend of magnetic parameters is consistent with our previous work,[27] though the improved accuracy has resulted due to the increased sintering time of the fresh samples prepared, with density of the pellets increased and the possibility of oxygen vacancies reduced.

The AFM transition is more obvious in the differentiated *dM/dT* curves, shown in field-cooled (FC) and zero-field-cooled (ZFC) cycles (right-panels, bottom/left axes). The transition temperatures $T_N$ (VSM precision ±10mK) obtained from the discontinuity in the slope is found to be 71K for $YMnO_3$, 71K for $YMn_{0.95}Fe_{0.05}O_3$, and 70K for $YMn_{0.9}Fe_{0.1}O_3$. Also shown (right-panels, top/right axes) is the relative difference ($M_{FC}/M_{ZFC}$-1) between the field-cooled warming (FCW) and zero-field cooled warming (ZFCW) data. This signifies weak ferromagnetism (WFM) evolution below $T_N$,[25,28] since the (otherwise fluctuating) field-aligned (FC) spins are increasingly stabilized in the ordered state. Due to higher magnetic frustration (*f*) implying larger AFM-correlations above $T_N$ in pure $YMnO_3$, the spontaneous internal field at $T_N$ is less robust,[29] reflected as less-prominent anomalies in both $1/\chi$-$T$ and *dM*/*dT* at its AFM-transition. Therefore, regarding the ability to field-align the fluctuating-spins across $T_N$ in an FC-protocol, external field reckons relatively stronger in pure versus the doped specimens, resulting in an observable FC-ZFC hysteresis near $T_N$ in the former. In contrast, Fe-doped specimens show large FC-ZFC hysteresis at *low temperatures*, versus a much-smaller one in the pure $YMnO_3$. Due to magnetic disorder, Fe-doped samples feature smaller saturation ordered moments ($\mu_{ord}$, obtained from the ND data and shown later in fig.8) vis-à-vis the pure $YMnO_3$. The *non* AFM-ordering Fe- and the *less* AFM-ordered Mn- spins (in the $\Gamma_1$ and/or $\Gamma_2$ domains; short-range FM Fe-O-Mn interactions[25] competing with long-range AFM Mn-O-Mn one), and the weakly-ordered spins residing in the $\Gamma_1$-$\Gamma_2$ domain-walls (as explained later), are all susceptible to field-align in an FC protocol, at low-temperatures. This may explain the large (smaller) low-*T* FC-ZFC hysteresis in the doped (pure) specimens. Note the higher saturation-value of the relative difference ($M_{FC}/M_{ZFC}$-1) for 5% Fe-doping, compared to both the pure and 10% Fe-doped compositions; later traced to the inhomogeneous magnetic structure of $YMn_{0.95}Fe_{0.05}O_3$.

Figure 3 shows the Rietveld-refined ND patterns of $YMn_{1-x}Fe_xO_3$ ($x$=0, 0.05, and 0.1) at 300K and 10K. The absence of any unaccounted peak in the refined data confirms the single-phasic nature of the specimens. Now, in order to determine the magnetic structure of our specimens, zero-field neutron diffraction data collected at 10K were used. Figure 4 shows the observed intensity-plots over a zoomed-in 2Θ-range at 10K for all three specimens. It is obvious from the low-temperature ND pattern that with Fe-doping at the Mn-site, the weightage of the prominent magnetic peak (100) at 2Θ~16.04° in the pure $YMnO_3$ gets systematically transferred to the (101) peak at 2Θ~17.67°. Also, the intensity of the purely magnetic (100) Bragg peak is found to decrease in $YMn_{0.95}Fe_{0.05}O_3$, while it completely disappears in $YMn_{0.9}Fe_{0.1}O_3$. Additionally, thermal evolutions of (100) and (101) peaks shown in figure 5 clearly evidence their intensity ratio $I_{(100)}/I_{(101)}$ varying with the Fe-content. These observations correspond to the change with Fe-doping, of Mn's ordered-magnetic-moment, and its orientation in the *a-b* plane. Raw-data manifestation of the magnetic structure modification vs. nominal doping is further confirmed by the magnetic phase analysis as follows.

Intensity of the magnetic peaks increases to constant values at 10K, suggesting that the magnetic structures have stabilized by the lowest temperature covered. In the attempt to fit the data, all 1D irreducible representation vectors $\Gamma_1$, $\Gamma_2$, $\Gamma_3$, and $\Gamma_4$ were generated for propagation vector $\boldsymbol{k} = 0$, by using BasIreps program in FULLPROF suite.[22] Figure 6 shows the Rietveld-refined neutron diffraction patterns at 10K without field (for all compositions) and under 5T-field (for the doped specimens). The data across 2θ~60°, due to the contribution of peaks from the shroud of the cryogen-free magnet, was excluded in the refinement process. Now, firstly considering the case of $YMnO_3$, we had tried to fit the data with various possible magnetic models proposed in the literature (as discussed in the introduction). The best fit though was obtained for the magnetic structure given by the basis vectors of the irreducible representation $\Gamma_1$ (fig.6(a)), consistent with the previous reports.[7,12,21] Moreover, the neutron diffraction data of $YMn_{0.95}Fe_{0.05}O_3$ was best fit with the admixture ($\Gamma_1+\Gamma_2$) of irreducible representations (fig. 6(b), (c)). Lastly, the $\Gamma_2$ representation was best fit for $YMn_{0.9}Fe_{0.1}O_3$ (figs. 6(d), (e)).

The change of magnetic structure on Fe-doping is but compositional analogue of the changes observed across the spin-reorientation temperature ($T_{sr}$) in e.g., $ScMnO_3$ and $HoMnO_3$,[7,17] and in $YMnO_3$ under pressure.[16] Therefore, here we maidenly witness the chemically-driven spin-reorientation by perturbative Fe-doping in $YMnO_3$. The mixed ($\Gamma_1+\Gamma_2$) configuration reflects inhomogeneous magnetic structure for the 5% Fe-doped $YMn_{0.95}Fe_{0.05}O_3$, with the possibility of statistically-distributed $\Gamma_1$ and $\Gamma_2$ 'domains'. Their interfacial "domain-walls spins" would exhibit higher propensity to harbour weak ferromagnetism (WFM), much the same way as found by breaking the long-range cycloidal-AFM order in $BiFeO_3$.[30] As seen earlier in our bulk magnetization results, this circumstance is in fact consistent with the higher saturation value of the relative change ($M_{FC}/M_{ZFC}$-1) for the mixed-configuration $YMn_{0.95}Fe_{0.05}O_3$, compared to the same for both the pure $YMnO_3$ ($\Gamma_1$) and $YMn_{0.9}Fe_{0.1}O_3$ ($\Gamma_2$) uniform magnetic structures. WFM may arise from fluctuating local-spins[29] that either (a) do not order antiferromagnetically (e.g., the Fe-spins in the present doped-specimens, or (b) are not robustly aligned at the transition temperature (due to a weaker spontaneous magnetization, traceable to spin-correlations in the disordered state), or (c) remain uncompensated at low-temperatures, due to the breakage of uniform magnetic structure by impurity (e.g., doping-induced disruption of the long-range cycloidal magnetic-order in $BiFeO_3$,[30] or of the triangular Mn-spins' arrangement by Fe in the present case), or (d) inhabit the 'interfacial' regions (weakly-ordered/uncompensated) between existing differently AFM-ordered 'domains' (such as $\Gamma_1$ and $\Gamma_2$ in the present $YMn_{0.95}Fe_{0.05}O_3$ specimen). This, as well as canting (absence of the exact cancellation of skewed-AFM-ordered moments[31]) produced either by an applied field (in the present case of ND at 5T), material-anisotropy, or short-range FM (viz., inverse D-M interaction)[25,28] all provide sources of weak ferromagnetism (WFM); producing split of the FC-ZFC magnetizations and manifesting as field-dependent reduced ordered-moments ($\mu_{ord}$, determined as 'AFM-aligned-projection' of the field-skewed moment).

Singh *et al.*[21] have shown that the ND data of pure $YMnO_3$ under an applied field of 5T field could be refined by using the same $\Gamma_1$ irreducible representation. Therefore, to further examine we have investigated the field-effect on the doped specimens only, by carrying out their neutron diffraction under 5T field. For this, after collecting the data at each temperature, field was reduced to zero and sample temperature was raised well beyond the Nèel temperature, to erase the possibility of frozen field-aligned spins. Figs. 6(c) & (e) show the Rietveld-refined in-field neutron diffraction patterns at 10K for $YMn_{0.95}Fe_{0.05}O_3$ and $YMn_{0.9}Fe_{0.1}O_3$ respectively. It is found that the ND data taken under 5T field could be refined by using the same irreducible representations as assigned for the zero-field data; ($\Gamma_1$+$\Gamma_2$) for $YMn_{0.95}Fe_{0.05}O_3$ and $\Gamma_2$ for $YMn_{0.9}Fe_{0.1}O_3$.

Figure7 (left panel) show temperature dependence of the lattice parameters *a* and *c* for $YMnO_3$, $YMn_{0.95}Fe_{0.05}O_3$, and $YMn_{0.9}Fe_{0.1}O_3$, with the field-induced alteration shown only for doped specimens, as obtained by the Rietveld refinement. The refined lattice parameters (under zero-field) match well with the previous reports.[9,18-19] The lattice parameter *a* (*c*) shows the positive (negative) thermal expansion. Thermal contraction of the lattice parameter *c* is reported to be persistent up to the ferroelectric transition.[9,32] Both *a* and *c* show large changes above $T_N$, and the applied field further alters the values of all the lattice parameters. In particular, note the *magneto-expansion* of *c* lattice constant for the doped-specimens here, as opposed to its *magneto-contraction* reported[21] for the pure specimen. Positive magneto-expansion of *c* reflects increased tilting (buckling) of MnO5 polyhedra (Y-O planes), indicating enhanced ferroelectric polarization *P* under the field, as against latter's decrease with *thermal contraction* of *c*, restoring the inversion-symmetry. The bond-lengths under zero and 5T field at 10K are summarized in TABLE I. Alteration of these bonds with the applied field confirms the presence of strong magneto-elastic (MEL) coupling in the doped specimens as well, though their observed trend here is different from those reported[21] for the pure $YMnO_3$. Figure 7 (right panel) shows the temperature-evolution and field-alteration of the planar bond-length (Mn-$O_p$, average of one Mn-O(3) and two Mn-O(4) equatorial bond-lengths), which governs the magnetic ordering, and is found to increase with temperature for all three compositions and also with field for the doped-specimens. Also, the slight increase in the planar bond-length with the Fe-content indicates the weakening of the superexchange interaction due to the dilution of Mn-ions and therefore, suppresses $T_N$ minutely. These results well correlate with our magnetization and specific heat data.

Figure 8 shows the *T*-dependence for pure & doped specimens and *H*-alteration for the latter, of the ordered magnetic moment ($\mu_{ord}$) at the Mn-site, explicitly obtained by the Rietveld refinement of our neutron diffraction data. Below $T_N$, $\mu_{ord}$ increases for all three compositions and reaches their respective saturation values. For pure $YMnO_3$, the saturation $\mu_{ord}$ is estimated as 3.41$\mu_B$; smaller than the ionic value of 4.0$\mu_B$ for $Mn^{+3}$. Therefore, about 0.6$\mu_B$ (15%) of the Mn-moment is still fluctuating at 10K in $YMnO_3$. This difference is the microscopic evidence of 2D geometrical frustration of the Mn-lattice. The saturation-value is 3.16$\mu_B$ for $YMn_{0.95}Fe_{0.05}O_3$ and 3.08$\mu_B$ for $YMn_{0.9}Fe_{0.1}O_3$; interestingly, the ratio 3.08/3.41 = 0.903 of $\mu_{ord}$(10K) for 10% Fe-doped and pure (undoped) specimens reflects but the compositional ratio (1-$x$) of their Mn-contents, as also confirmed (shown below) up to 60K. This precise scaling leads to an important insight that the chemo-magnetically disordered Fe-dopant's moment is *fully-fluctuating* and thus non-contributing to the measured ordered-moment. The nearly parallel lines drawn at $T < T_N$ in Fig.8-inset, connecting the [scaled by $\mu_{ord}$(10K, $x$=0)] zero-field ordered-moments of the two pure ($\Gamma_1$/$YMnO_3$ and $\Gamma_2$/$YMn_{0.9}Fe_{0.1}O_3$) structures illustrate the indifference (to $\Gamma_1$ or $\Gamma_2$) of individual Mn's ordered-moment part; solely contributing below $T_N$ to $\mu_{ord}(T)$, in proportion to their Mn-population alone. Moreover, clearly discernible *below-the-line* values of the scaled-$\mu_{ord}(T)$ for the mixed [($\Gamma_1$+$\Gamma_2$)/$YMn_{0.95}Fe_{0.05}O_3$] configuration reflect enhanced Mn-moments' fluctuation at 10K (≈ 17% here vs. ~15% in $YMnO_3$).

Under 5T field, saturation-$\mu_{ord}$ reduces further; to 95% for $YMn0_{0.95}Fe_{0.05}O_3$ and to 98% for $YMn_{0.9}Fe_{0.1}O_3$, of their respective zero-field values. Considering the applied field to cause off-plane canting (skewing) of the Mn-moments (as the ND-estimated ordered-moment is but its average planar component, partaking in triangular AFM-formation), the (vertical) canting-angle would therefore satisfy $\cos\psi = \mu_{ord}(5T)/\mu_{ord}(0T)$. Angles $\psi$ thus evaluated at 10K/70K is ≈10.58°/19.36° for $YMn_{0.95}Fe_{0.05}O_3$ and ≈10.13°/28.56° for $YMn_{0.9}Fe_{0.1}O_3$. Evidently, the more frustrated specimen exhibits less susceptibility to canting near $T_N$, albeit being almost equally susceptible at low $T$'s. Anti-regression of the *maximum* (70K) vertical-canting and planar-frustration of the moments is beautifully underscored by the excellent equality of the ratio $[\cos\psi_{5\%Fe}/\cos\psi_{10\%Fe}]_{70K}$ =1.074 of moment's canting-direction-cosines, and that of the bulk frustrations $f_{5\%Fe}/f_{10\%Fe}$ =1.072, for $YMn_{0.95}Fe_{0.05}O_3$ and $YMn_{0.9}Fe_{0.1}O_3$ specimens. Moreover, at low-$T$'s we similarly note that the compositional-ratio of the (5T) field-induced canted-WFM moments (C-WFM) $[(\mu_{ord}\sin\psi)_{5\%Fe}/(\mu_{ord}\sin\psi)_{10\%Fe}]_{10K}$ =1.072 agrees (to ≤ 10%) with that of the measures of their saturation weak ferromagnetism (S-WFM), definable as $[(M_{FC}/M_{ZFC}\text{-}1)_{5\%Fe}/(M_{FC}/M_{ZFC}\text{-}1)_{10\%Fe}]_{10K}$ =1.176. The larger difference here being due to the direct origin (post ZFC, 5T-field in the AFM state) of the former and the quenched-canted origin (post FC, 100 Oe field in the PM state) of the latter ratio.

To further examine the more interesting *physically*-mixed $YMn_{0.95}Fe_{0.05}O_3$ composition, *mathematical* $\Phi(T)$ evaluated as the 'statistical-average' of the (zero-field) angle between the Mn-moment and the $a$-axis (90° for $\Gamma_1$ and 0° for $\Gamma_2$ configuration[7,12]) is shown in fig.9 (left $y$-axis). This lies between 77.98° (10K) to 3.56° (70K); providing the base-data for thermal evolution of the mixed ($\Gamma_1$+$\Gamma_2$) magnetic structure in $YMn_{0.95}Fe_{0.05}O_3$. Now, the relative abundance $p_1$ ($p_2$=1-$p_1$) of $\Gamma_1$ ($\Gamma_2$) configuration satisfy $\Phi = 90p_1+0p_2$, so that $p_1= \Phi/90$ and $p_2=1-\Phi/90$. Moreover, referred to fig.8-inset, for this specimen the (ideal) *fully*-ordered-moment per Mn-site ought to satisfy $\mu_{ord}^{full}(x=0.05)/\mu(\mathrm{Mn}^{+3}) \approx 95\%$, with the ionic $\mu(\mathrm{Mn}^{+3})$ = 4.0$\mu_B$, whilst its (practical) *maximally*-ordered-moment is expected as $\mu_{ord}^{max}(T, x=0.05) \approx 0.95\mu_{ord}(T, x=0)$ at finite *T*'s. The low-*T*-dominant $\Gamma_1$-share (= $p_1[\mu_{ord}(T, x=0.05)/\mu_{ord}^{full}(x=0.05)]_{0T}$) is seen in fig.9 (right *y*-axis) to drop from 72% (10K) through 10% (60K) to 1.6% (70K), whereas the $\Gamma_2$-fraction (= $p_2[\mu_{ord}(T, x=0.05)/\mu_{ord}^{full}(x=0.05)]_{0T}$) rises from 11% (10K) through (max.) 52% (60K) up to 39% (70K) near $T_N$. The shortage of these fractions' sum below unity ($1-[\mu_{ord}(T, x=0.05)/\mu_{ord}^{full}(x=0.05)]_{0T}$) represents the gross-fraction of residual-fluctuating Mn-spins; its temperature-evolution (~17% at 10K to ~59% at 70K) shown in fig.9-inset (left *y*-axis). Furthermore, estimated net-fraction of weakly-ordered Mn-spins in the interfacial $\Gamma_1$-$\Gamma_2$ domain-walls ($1-[\mu_{ord}(T, x=0.05)/\mu_{ord}^{max}(T, x=0.05)]_{0T}$) reduces from (max.) 2.2% at 10K (fig.9-inset, right *y*-axis). These results are consistent with our $\mu_{ord}$ and Φ-angle analyses.

## CONCLUSIONS

In summary, we have carried out neutron diffraction study of perturbatively (up to 10%) Fe-doped polycrystalline *h*-$YMn_{1-x}Fe_xO_3$ below and above the Nèel temperature in zero- (all specimens) and under 5T-field (doped compositions). The lattice parameters and planar bond-lengths show large (minute) changes above (below) $T_N$, nearly uniformly altered (in the doped specimens) by the 5T field across the AFM and PM states, exhibiting strong magneto-elastic nature of the system. Our study highlights spin re-orientation in the hexagonal manganite, composition-driven by the magnetic (Fe) doping; the magnetic ground state changes from a highly-frustrated ($\Gamma_1$/$YMnO_3$) to a lowly-frustrated ($\Gamma_2$/$YMn_{0.9}Fe_{0.1}O_3$) magnetic structure, via a mixed [($\Gamma_1$+$\Gamma_2$)/$YMn_{0.95}Fe_{0.05}O_3$] configuration, which thermally evolves from $\Gamma_1$-rich (low-*T*) to $\Gamma_2$-rich (near-$T_N$) character. The applied 5T field does not qualitatively alter the overall magnetic structures of the doped specimens; it adds to the disorder-induced suppression of the evaluated/planar-component of the ordered moment ($\mu_{ord}$) via latter's off-plane/vertical canting. Therefore, magnetic field enhances the doping-induced relief of the planar magnetic frustration. Concerning geometric multiferroicity of the system, a qualitative change in its magneto-electric coupling hand-in-hand with disorder-driven spin-reorientation emerges from the present study. This important implication is compelled by the observed magneto-expansion of the *c* lattice constant (signifying[9] +ve magneto-polarisation $dP/dH$) in the doped specimens, vis-à-vis literature-reported[21] magneto-compression of *c* (i.e., –ve $dP/dH$) in pure $YMnO_3$.

## ACKNOWLEDGEMENTS

We thank Mukul Gupta, V. Ganesan, and A. Banerjee for providing XRD, specific heat, and magnetization facilities respectively, and D. Kumar for help with the magnetization measurements. P. Chaddah is acknowledged for his scientific support and encouragement.

## Table Captions

**Table-I**. Mn-O [O(1), O(2), O(3), and O(4)] bond-lengths and Mn--Mn distance at 10K for $YMnO_3$ (0T), $YMn_{0.95}Fe_{0.05}O_3$ (0T, 5T), and $YMn_{0.9}Fe_{0.1}O_3$ (0T, 5T).

**TABLE I**

| Specimen / Bond-Length (Å) | $YMnO_3$ (0T) | $YMn_{0.95}Fe_{0.05}O_3$ (0T) | $YMn_{0.95}Fe_{0.05}O_3$ (5T) | $YMn_{0.9}Fe_{0.1}O_3$ (0T) | $YMn_{0.9}Fe_{0.1}O_3$ (5T) |
|---|---|---|---|---|---|
| **Mn-O(1)** | 1.75491(8) | 1.75771(9) | 1.75756(9) | 1.75792(8) | 1.75840(9) |
| **Mn-O(2)** | 1.97229(9) | 1.97542(10) | 1.97525(10) | 1.97566(9) | 1.97620(10) |
| **Mn-O(3)** | 2.00750(5) | 2.00859(10) | 2.00904(6) | 2.00904(5) | 2.00930(5) |
| **Mn-O(4)** | 2.07165(4) | 2.07274(7) | 2.07322(5) | 2.07321(4) | 2.07348(4) |
| **Mn--Mn** | 3.57985(10) | 3.58217(9) | 3.58254(10) | 3.58272(9) | 3.58286(8) |

## Figure Captions

**Fig.1.** The temperature dependence of $C_p/T$ for $YMn_{1-x}Fe_xO_3$ ($x$= 0, 0.05, and 0.1) under zero- and 5T-fields. Open (filled) symbols show the data points in zero (5T) field.

**Fig.2.** Left panels: inverse-susceptibility ($1/\chi = H/M$) vs. temperature plot for $YMn_{1-x}Fe_xO_3$ ($x$= 0, 0.05, and 0.1). Straight lines (fitted over 150-300K data) correspond to the Curie-Weiss law. Right panels: FC-ZFC magnetization hysteresis, clearly manifested in the derivatives $dM/dT$ (bottom/left axes) and the relative change between the FCW/ZFCW data (top/right axes), as a quantitative measure of the weak ferromagnetism (WFM) evolution in the ordered AFM state, with different prominence-regimes in pure (around $T_N$) and doped (low-$T$'s) specimens, and its predominance in the mixed ($\Gamma_1+\Gamma_2$) magnetic-structured $YMn_{0.95}Fe_{0.5}O_3$.

**Fig.3.** (a), (c), &( e)-- Rietveld-refined neutron diffraction pattern for $YMn_{1-x}Fe_xO_3$ ($x$= 0, 0.05, and 0.1), taken at 300K. (b), (d), & (f)-- Rietveld-refined neutron diffraction pattern for $YMn_{1-x}Fe_xO_3$ ($x$= 0, 0.05, and 0.1) taken at 10K under zero field.

**Fig.4.** Bragg-peak pattern over zoomed-in 2Θ-range for $YMn_{1-x}Fe_xO_3$ ($x$= 0, 0.05, and 0.1), taken at 10K under zero field. The data are shifted upwards for clarity. Inset shows the survey pattern for $YMnO_3$.

**Fig.5.** Intensity vs. temperature plot of (100) & (101) magnetic peaks at zero field for $YMn_{1-x}Fe_xO_3$ ($x$= 0, 0.05, and 0.1).

**Fig.6.** Observed (10K) and calculated neutron diffraction patterns with $P6_3cm$ symmetry for $YMn_{1-x}Fe_xO_3$ ($x$= 0, 0.05, and 0.1). To refine the data, we used (a) $\Gamma_1$ representation for $YMnO_3$, (b) & (c) ($\Gamma_1+\Gamma_2$) representation for $YMn_{0.95}Fe_{0.05}O_3$ under zero- and 5T-fields respectively, and (d) & (e) $\Gamma_2$-representation for $YMn_{0.9}Fe_{0.1}O_3$ under zero- and 5T-fields, respectively. Lines below the bars indicate the difference between the observed and calculated neutron diffraction patterns. Upper and lower bars correspond respectively to nuclear and magnetic Bragg peaks.

**Fig.7.** Left panels: temperature-dependence of the lattice parameters $a$ and $c$ for $YMn_{1-x}Fe_xO_3$ (all compositions) and their alteration under 5T-field (doped-specimens). Right panels: average planar (Mn-$O_p$) bond-length evaluated by using Mn-O(3) & Mn-O(4) bond-lengths under zero-field (open symbols) for $YMn_{1-x}Fe_xO_3$ (all compositions) and under 5T-field (closed symbols) for the doped-specimens. Insets depict the variations of the cell-volume.

**Fig.8.** Temperature-dependence of the ordered magnetic moment $\mu_{ord}$ in zero-field (open symbols) for $YMn_{1-x}Fe_xO_3$ (all compositions), and under 5T-field (closed symbols) for the doped-specimens. Inset depicts (scaled to $\mu_{ord}$(10K) of the pure $YMnO_3$) zero-field $\mu_{ord}$ vs. the Fe-content at 10K, 30K, 50K, and 60K; evidencing nil contribution from the fully-fluctuating Fe-moments [$\mu_{ord}(x=0.1) \equiv 0.9\mu_{ord}(x=0)$], and smaller than maximum-expected values for the mixed-structured $YMn_{0.95}Fe_{0.05}O_3$ [$\mu_{ord} < \mu_{ord}^{max} \approx 0.95\mu_{ord}(T, x=0)$], due to the weakly-ordered Mn-moments trapped in the $\Gamma_1$-$\Gamma_2$ interfacial domain walls.

**Fig.9.** Left $y$-axis: zero-field $\Phi(T)$ as average over abundance, of the angle (90° for $\Gamma_1$ and 0° for $\Gamma_2$) between Mn-moment and the $a$-axis, in the mixed-structured $YMn_{0.95}Fe_{0.05}O_3$. Right $y$-axis: consequent $T$-dependences of $\Gamma_1$- and $\Gamma_2$-fractions; the magnetic structure evolves from $\Gamma_1$-rich (low-$T$) to $\Gamma_2$-rich (near-$T_N$) configuration. Inset: left $y$-axis-- thermal-evolution of the gross-fraction of residual-fluctuating Mn-moments (1-[$\mu_{ord}(T, x=0.05)/\mu_{ord}^{full}(x=0.05)$]); right $y$-axis-- thermal-dissolution of estimated net-fraction of weakly-ordered Mn-moments trapped within the $\Gamma_1$-$\Gamma_2$ domain walls (1-[$\mu_{ord}(T, x=0.05)/\mu_{ord}^{max}(T, x=0.05)$]).

**Fig.1**

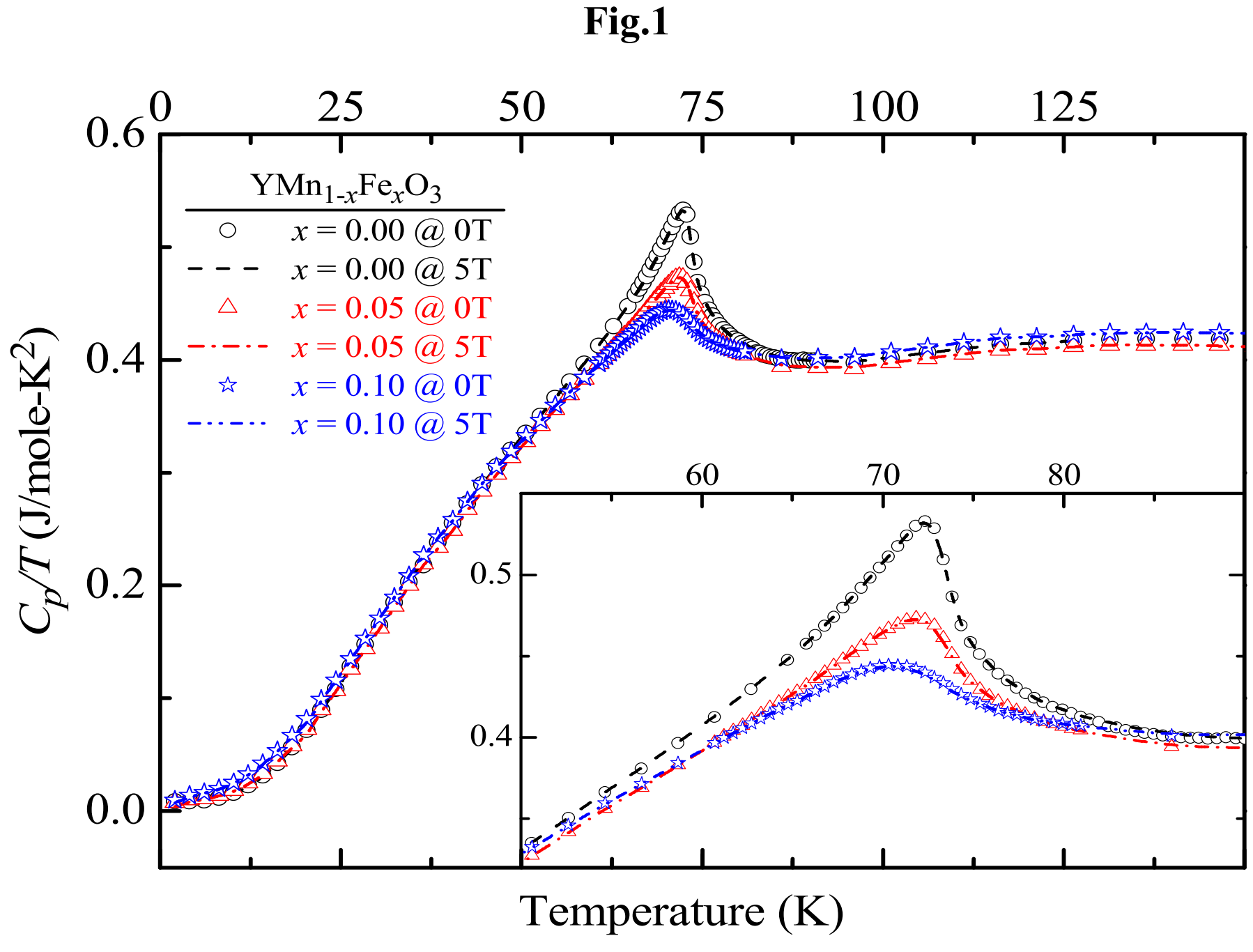

**Fig.2**

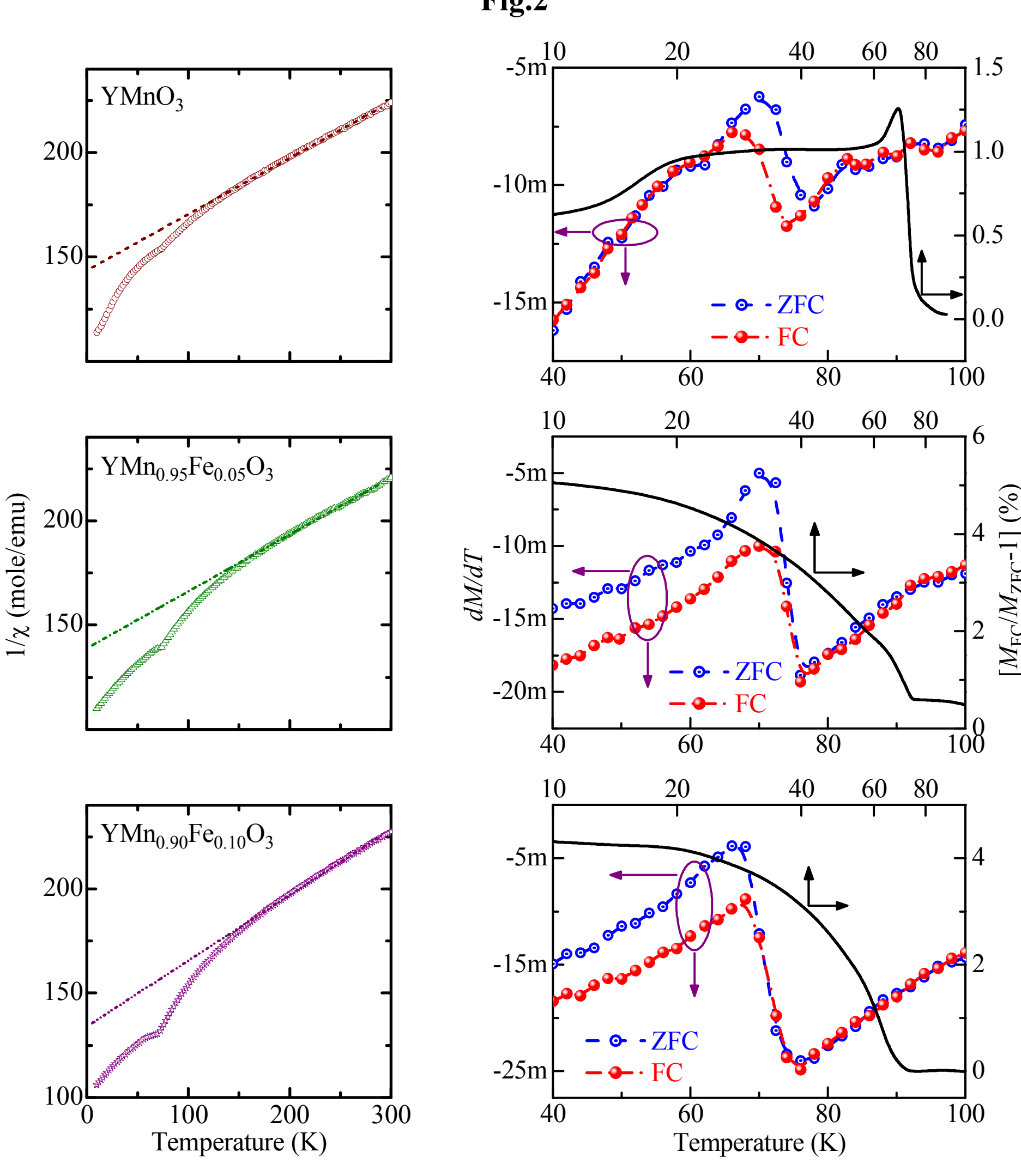

**Fig.3**

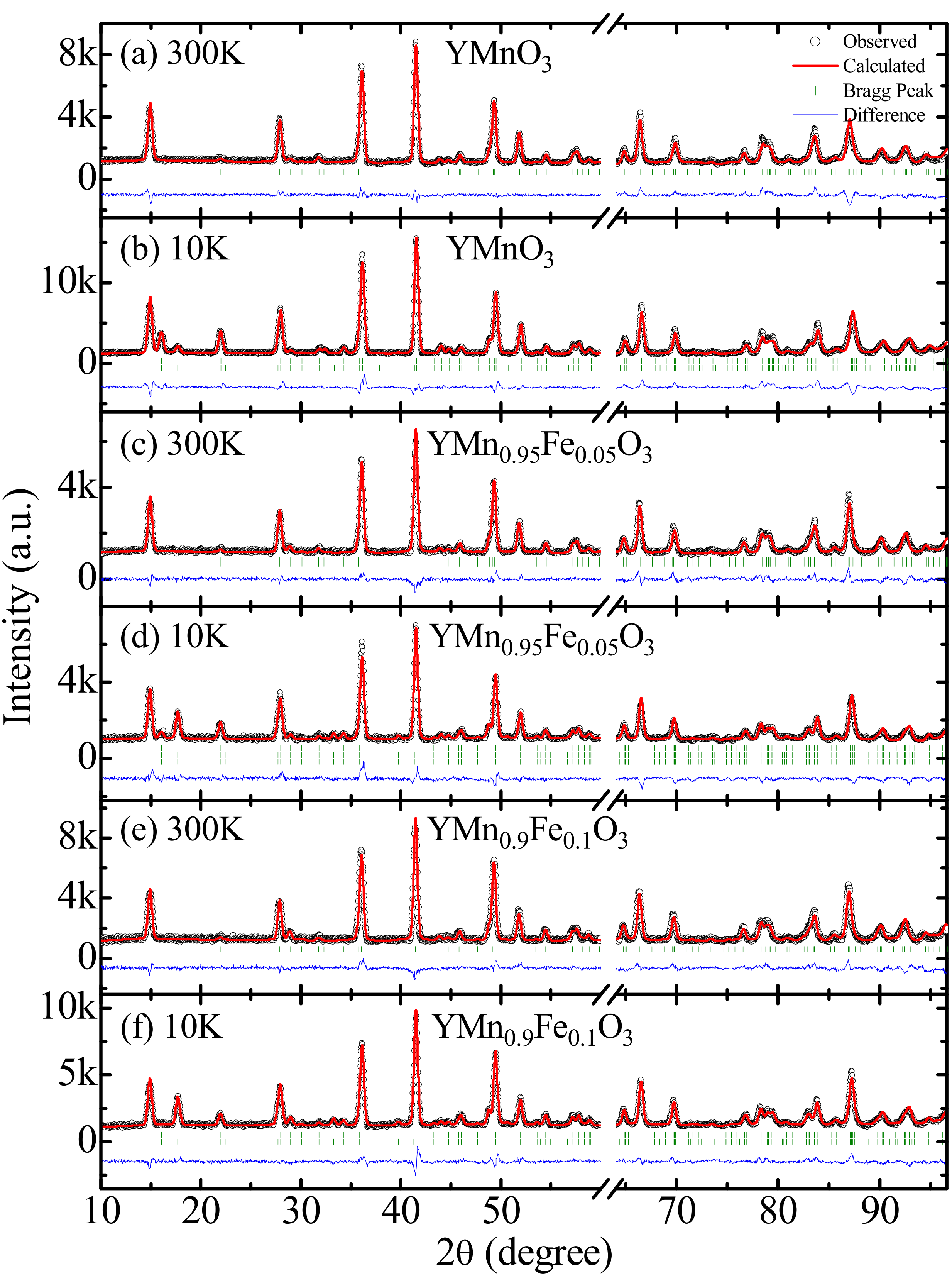

**Fig.4**

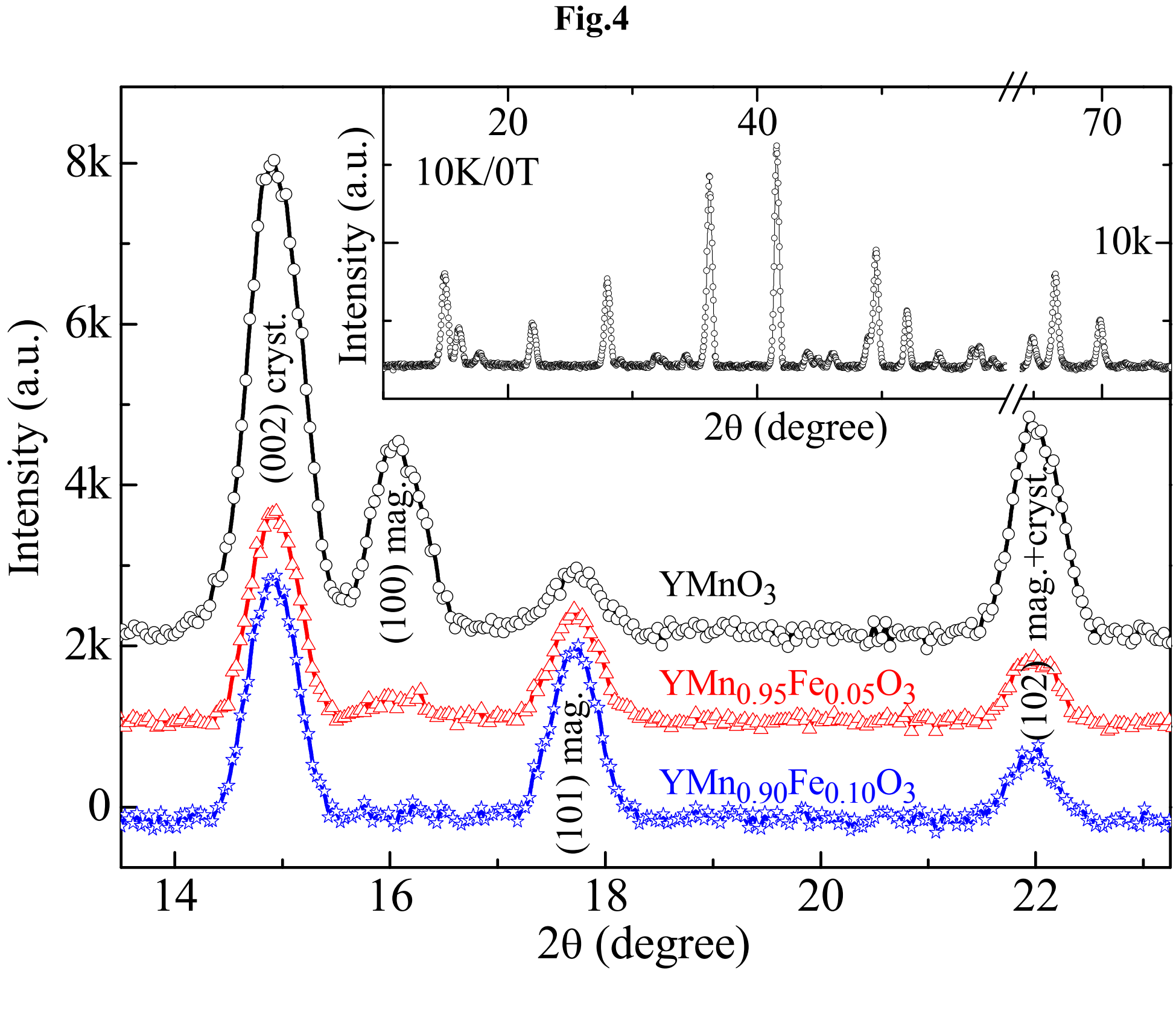

**Fig.5**

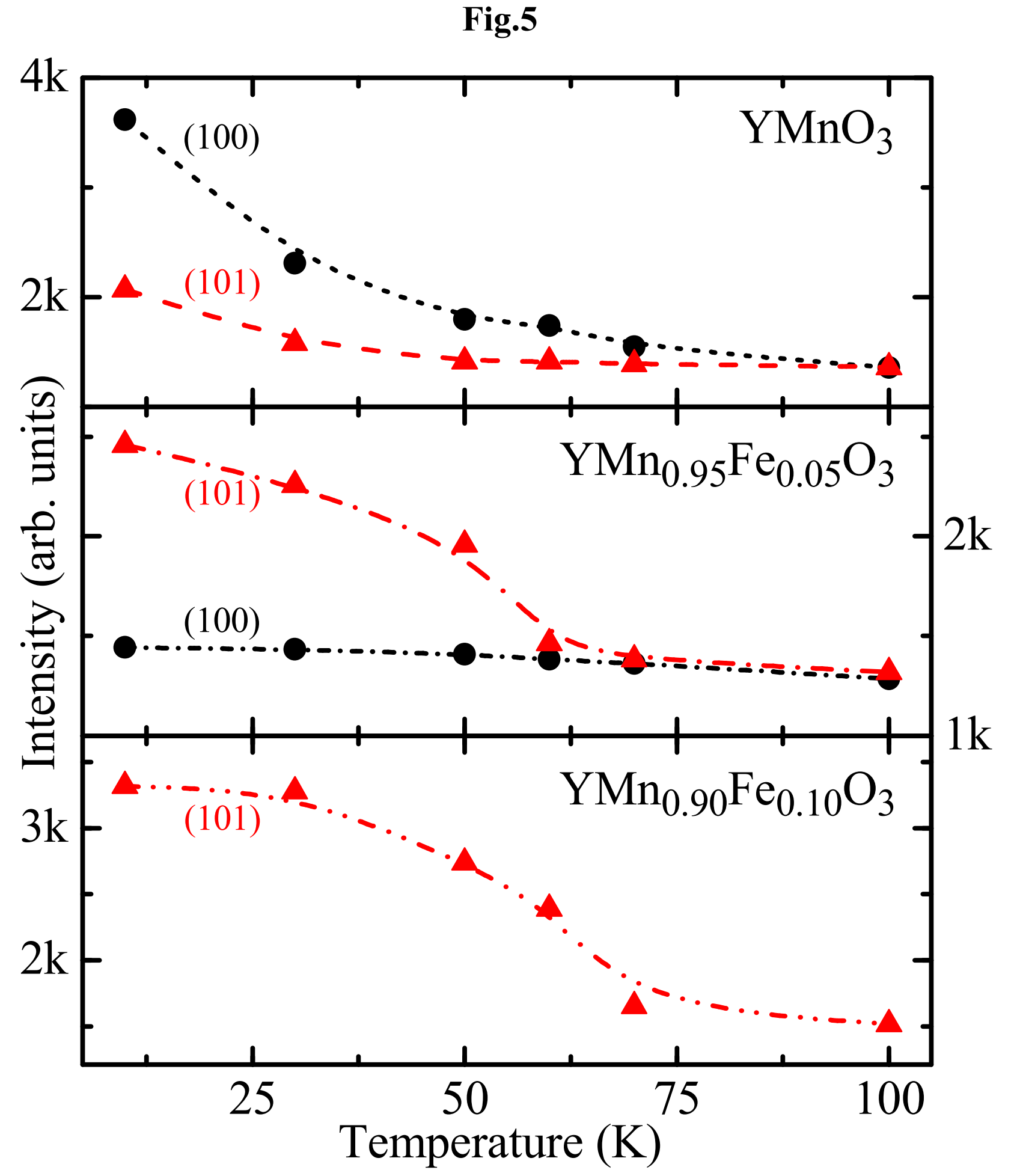

**Fig.6**

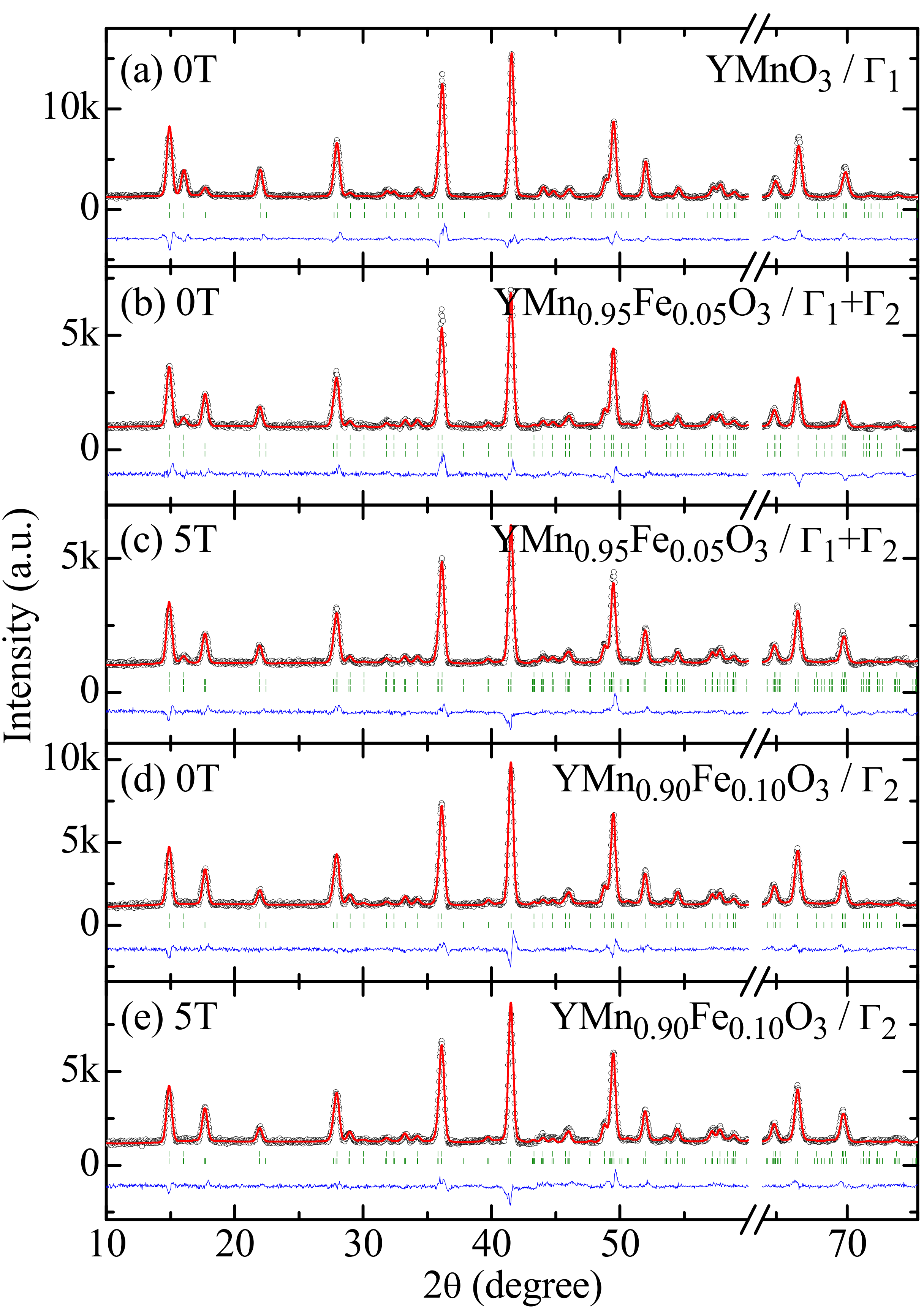

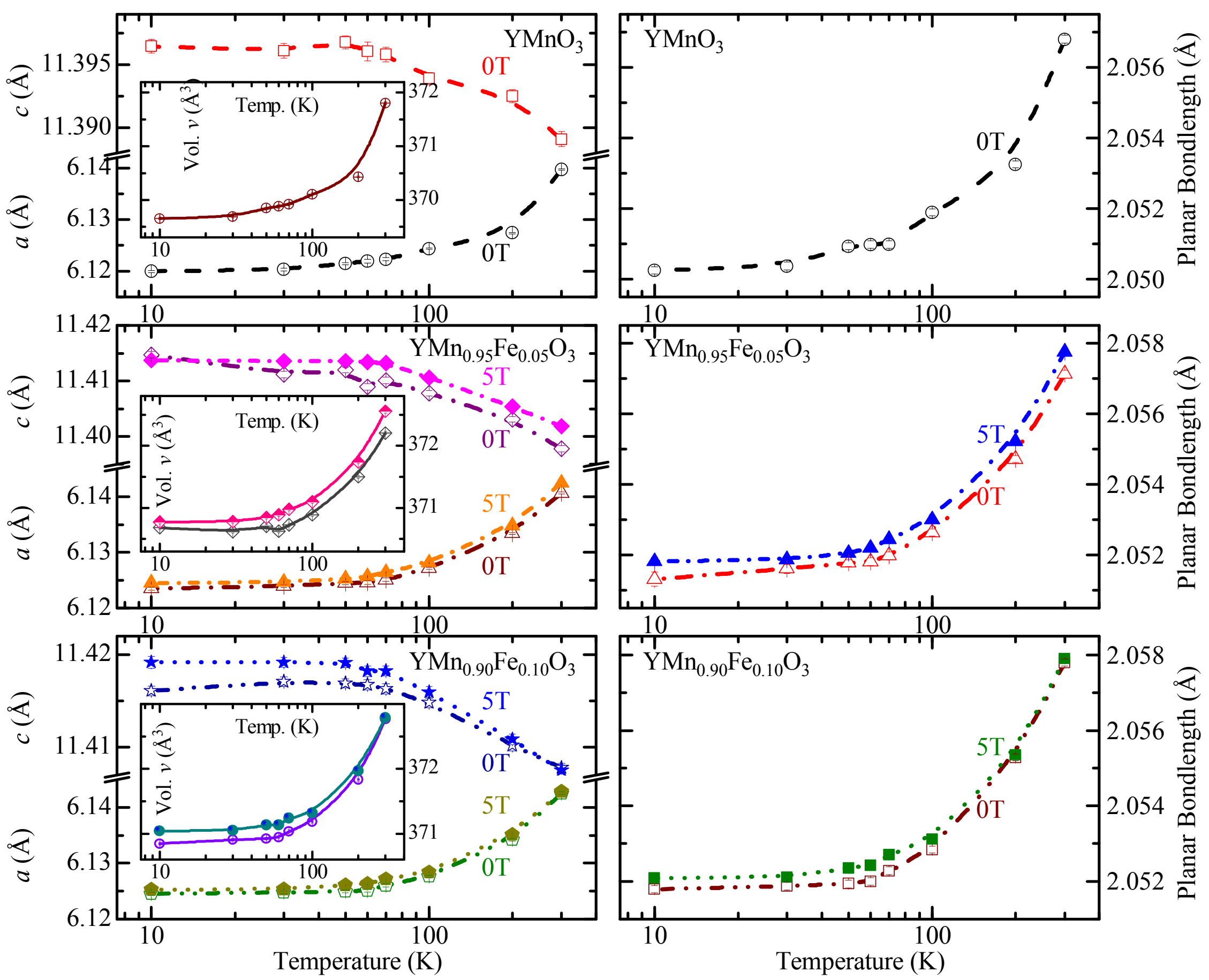

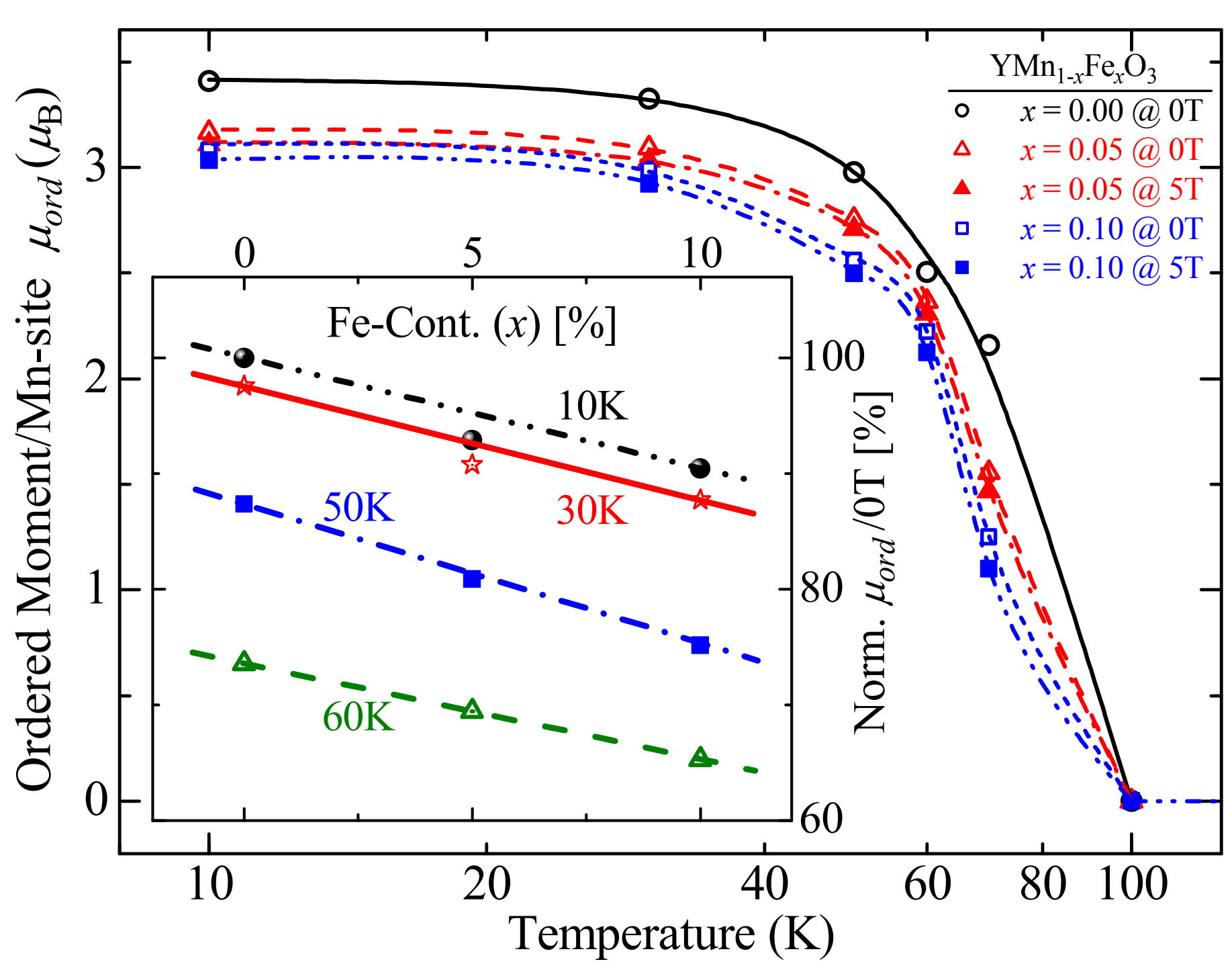

**Fig.9**

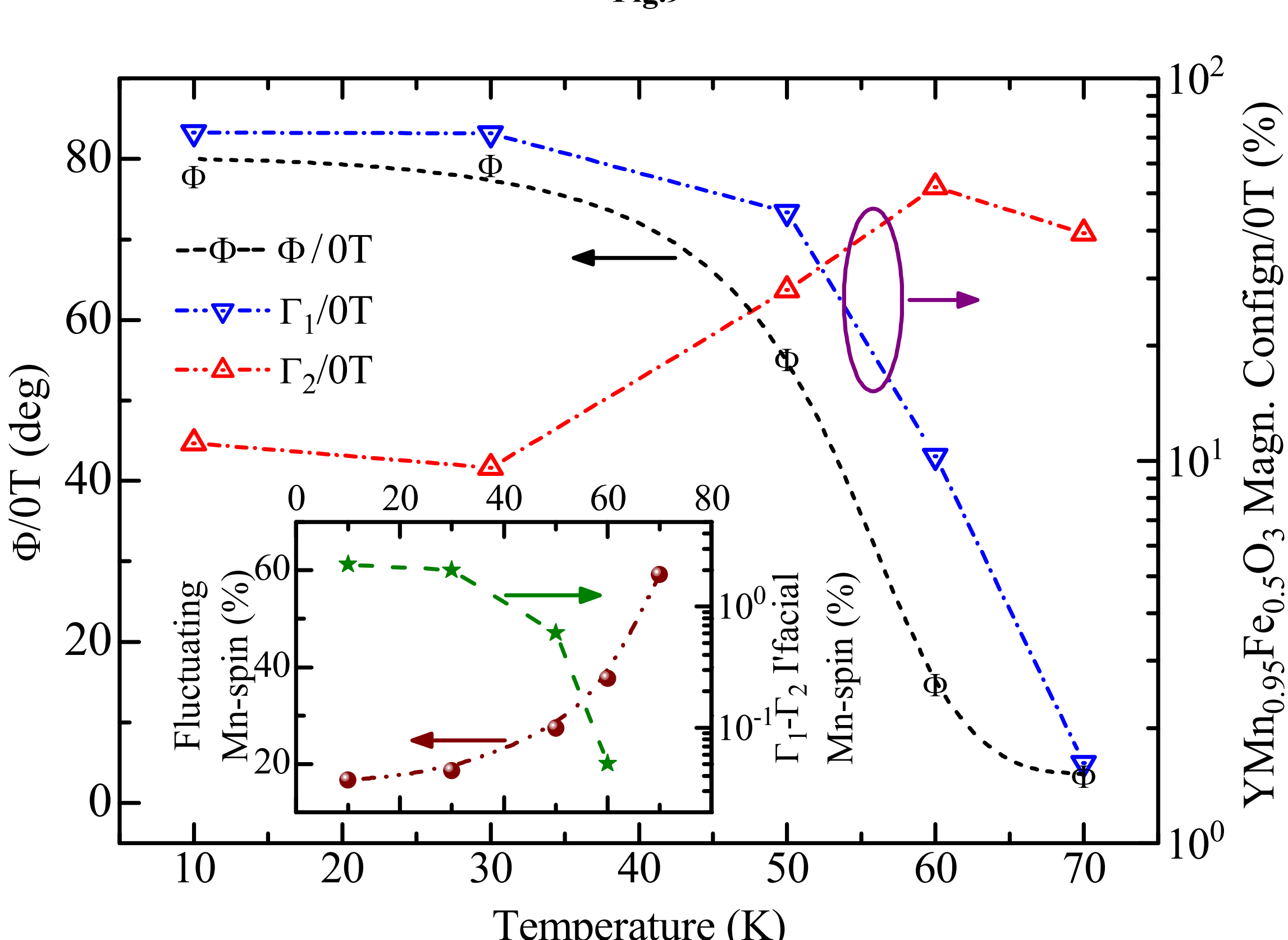